\documentclass{elsart}
\usepackage{natbib,epsf}

\def\beq{\begin{equation}}
\def\eeq{\end{equation}}

\begin{document}
\sloppy

\begin{frontmatter}

\title      {Parametric, nonparametric and parametric modelling 
of a chaotic circuit time series}

\author   {J.~Timmer\thanksref {corr}}
\author   {H.~Rust} 
\author   {W.~Horbelt} 
\author   {H.U.~Voss} 
\address {Freiburger Zentrum f\"ur Datenanalyse und Modellbildung,
                Eckerstr. 1, D-79104 Freiburg, Germany}
\thanks[corr] {Corresponding author. Tel.: +49/761/203-5829, 
FAX:+49/761/203-5967 ; e-mail: jeti@fdm.uni-freiburg.de} 

\date {\today }
\maketitle

\begin {abstract}

The determination of a differential equation underlying
a measured time series is a frequently arising task in
nonlinear time series analysis. In the validation of a proposed 
model one often faces the dilemma that it is
hard to decide whether possible discrepancies between  the 
time series and model output are caused
by an inappropriate model or by 
bad estimates of parameters in a correct type of model, or both.
We propose a combination of parametric modelling 
based on Bock's multiple shooting algorithm and nonparametric modelling
based on optimal transformations as a strategy to
test proposed models and if rejected suggest and test new ones.
We exemplify this strategy on an experimental time series from
a chaotic circuit where we obtain an extremely accurate reconstruction
of the observed attractor. 

\end {abstract}

\begin {keyword}
 Modelling chaotic time series; nonparametric modelling; parameter estimation;
 flow vector field reconstruction.
\end {keyword}
\end{frontmatter}

\section {Introduction} \label{intro}

It is an old goal in nonlinear time series analysis to infer 
the ``Equations of motion from data series''~\cite{crutchfield87}.
Especially, for continuous flow systems 
modelling a sampled time series by a differential
equation might allow for insight into the mechanisms at work
by interpreting the resulting structure of the equation and values of the 
parameters.
This is known as ``interpretability'' of a model~\cite{Elder96} as opposed 
to black-box approaches like an attractor reconstruction~\cite{packard80}.

The straightforward procedure to estimate the parameters in differential
equations is to estimate time derivatives 
from the data and determine the parameters by a least-squares
minimisation~\cite{cremers87,gouesbet96,hegger98}. 
This approach is firstly limited by the 
additive observational noise which usually covers the 
observed dynamics and prohibits the reliable estimation
of the derivatives~\cite{irving97}, especially if only one
component of the multidimensional system can be measured. 
Secondly, it is hampered by the huge number of possible nonlinear 
models that have to be compared. 

Fortunately, there is often prior knowledge that gives constraints on
the model or even suggests a specific type of model~\cite{rulkov94,weiss95}.
The validity of a specific 
model can be evaluated by comparing properties of simulated time series
with the measured one~\cite{weiss95}. This approach faces a dilemma.
For the simulation the parameters have to be specified. 
Thus, it is difficult to decide whether possible discrepancies 
between the properties of the simulated and the measured time series
are caused by the fact that the chosen type or structure of model is wrong or 
by the circumstance that the parameters have been chosen inappropriately in the
correct type of model, or both. In this paper we show that parametric modelling
based on Bock's multiple shooting algorithm~\cite{bock81,bock83} can solve 
this dilemma. If the model type is rejected we propose 
a search in structure space instead of parameter space.
A search in structure space can be performed in two ways:
Fitting coefficients for certain basis functions (e.g. coefficients of 
polynomials) and nonparametrically. Here we chose nonparametric  
modelling based on optimal transformations and maximal correlations 
\cite{breiman85,haerdle,voss97} as an
exploratory tool to suggest a new type of model that again
can be tested by parametric modelling.
We explain and exemplify this strategy on a measured time series from
a chaotic circuit~\cite{rulkov94,rulkov92}.
In this application our strategy yields a reconstruction of the 
observed attractor of unprecedented accuracy.

The paper is organized as followed: In the next section we briefly
describe the two modelling procedures. For detailed discussions of the
mathematical methods, proofs of convergence and numerical details, see
\cite{bock81,bock83,breiman85}. In Section  \ref{sec_data}
we introduce the data and the model derived from prior knowledge 
for these data in~\cite{rulkov94}. 
The parametric modelling based on the suggested
model, the search for a better model by a nonparametric procedure and the final
parametric fit is presented in Section \ref{results}.

\section{Methods} \label{methods}

\subsection{Bock's multiple shooting algorithm for parameter estimation} 
                \label{methods_bock}

A common setting in modelling time series by differential equations is
\begin{eqnarray}
  \dot{\vec{x}}(t) & = & \vec{f}(\vec{x}(t),\vec{p})\qquad 
  (\vec{x}(t)\in  R^m, \quad\vec{p}\in R^p, \quad t\in [t_0,t_f])\label {dyn_eq} \\
  \vec{y}(t_i) & = & \vec{g}(\vec{x}(t_i)) + \vec{\eta}(t_i) \label {obs_eq}\;,
\end{eqnarray}
where $\vec{f}$ defines the dynamics that depends on a parameter 
vector $\vec{p}$.
The state vector $\vec{x}(t)$
is observed through a function $\vec{g}(.)$ and the observation 
$\vec{y}(t_i)$ is sampled at times $t_i$ and disturbed by white observational
noise $\vec{\eta}(t_i)$ of standard deviation $\vec{\sigma}_i$. 
In general, the observation function $\vec{g}(.)$ can also contain unknown
parameters. In the following, for ease of clarity, we assume the often
met condition of a known, scalar observation function 

\beq \label{obs_function}
g(.) = x_1(t_i)
\eeq 
which records one, for ease of notation the first, component of the dynamical state. 

A first approach to estimate the parameters without the need to
estimate derivatives from the data is the {\it initial value approach}
\cite{edsberg95,schittkowski94}.
For this procedure, initial guesses for the parameters and
the initial values $\vec{x}(t_0)$ are chosen. Then Eq.~(\ref{dyn_eq})
is solved numerically and estimates $\hat{\vec{y}}(t_i,\vec{p},\hat{\vec{x}}(t_0))$ 
are calculated by Eq.~(\ref{obs_function}). The error cost function

\beq \label{error_fcn}
   \chi^2 = \frac{1}{N} \sum_{i=1}^{N} 
\frac{(\hat{y}(t_i,\vec{p},\hat{\vec{x}}(t_0)) - y(t_i))^2}{\sigma_i^2}
\eeq

is minimized with respect to the parameters $\vec{p}$ and the initial 
values $\vec{x}(t_0)$
by some numerical optimization algorithm~\cite{recipes}.
For this procedure, the only information from the measured time series
that enters the initial guesses of the optimization procedure is the 
value of the observed component, Eq.~(\ref{obs_function}), at time $t_0$.

Simulation studies have shown that, for many types of dynamics, 
this approach is numerically
unstable by yielding a diverging trajectory or stopping in a local minimum
~\cite{richter92,timmer98c,timmer00e}. The reason for this is that 
even for slightly wrong parameters, the trial trajectory 
looses contact to the measured trajectory. 
This is most evident in the case of chaotic dynamics, where due to the 
sensitivity with respect to initial conditions the
numerical trial trajectory is expected to follow the measured 
trajectory of the system only for a limited amount of time.
This divergence of the numerical and measured trajectory introduces many
local minima in the landscape of the error functional, Eq.~(\ref{error_fcn}).

This problem can be circumvented by a multiple shooting algorithm
introduced by Bock~\cite{bock81,bock83}.
Here we only briefly explain this algorithm.
The basic idea of the algorithm is to start the optimization 
with an only piecewise continuous trajectory that stays close to the data. 
If the 
observation function is $g(.)=x_1(t)$,
more information than only the first value of the measured time
series as in the initial value approach can be used as initial guesses for 
the optimization procedure by the following 
strategy: The time interval $[t_0,t_f]$ of measurement is divided into numerous 
segments $[t_j, t_{j+1}]$. A trial trajectory for each segment is
calculated using the information $\hat{x}_1(t_j)= y(t_j)$ from the measured
time series and initial
guesses for the remaining components of $\vec{x}(t_j)$.
The condition that the underlying trajectory is smooth 
enters into the algorithm by a constraint in the
cost function, Eq.~(\ref{error_fcn}). This constraint is nonlinear in 
the parameters but enters the optimization strategy using a Gau\ss{}-Newton 
procedure only in a linearized way. Therefore,
the trajectory is allowed to be discontinuous at the beginning of
the optimizing iterations but is forced to become smooth in the end.
After convergence the algorithm also provides an estimate 
 $\hat{\vec{x}}(t)$ of the unobserved dynamical state.

For time series of chaotic systems it will, in general, not be possible
to find a trajectory $\hat{\vec{x}}(t)$ that shadows the 
true trajectory for arbitrary long times. 
Furthermore, if the model is not correct, experience shows that a fit
to the whole time interval $[t_0,t_f]$ of measurement often does not converge.
For both reasons, the
time series is cut into pieces of equal length,
and the parameters are fitted to all the pieces simultaneously.
The length of these pieces is chosen as large as possible.

We show the process of convergence for the time series under investigation 
in Section \ref{resultsIII}, Fig.~\ref{second_fit}.
Analogous illustrative applications to simulated time series are given 
in~\cite{timmer00e,baake92a,timmer98d}.
Note that after convergence the algorithm is identical to 
an initial value approach: It predicts the time series for each of
the whole pieces based on the estimated initial values of
the unobserved state vector.

After convergence confidence intervals for the parameters can be
calculated from the curvature of the cost function~\cite{bock83}.
Note that both approaches do not need an attractor reconstruction by 
a delay embedding. 
Thus, all problems associated with the delay reconstruction of a chaotic and 
noisy phase space, like finding an optimal embedding window and
the ``curse of dimensionality,'' are of no or minor importance here.
Most importantly, we do not need such huge amounts of data as 
generally needed for a useful delay reconstruction and the method is
also applicable to transient time series, see~\cite{timmer98c,baake92b}.

By Bock's multiple shooting algorithm the probability of stopping in local
minima is reduced compared to the initial value approach.
Nonetheless, the algorithm should be applied with different
initial guesses for the parameters and the unobserved components 
at times $t_j$ to yield confidence
in the global optimality of the resulting estimates.

\subsection {Nonlinear regression and optimal transformations}
\label{methods_opt_trans}

Optimal transformations and the associated concept of 
maximal correlation provide a nonparametric
procedure to detect and determine nonlinear relationships in data sets. 
Let $X$ and $Y$ denote two zero-mean data sets and

\beq
R(X,Y) =\frac{E[XY]}{\sqrt{E[X^2]\,E[Y^2]}}
\eeq

their (normalized) linear correlation coefficient, where $E[.]$ is the 
expectation value. The basic idea of this approach is to find
transformations $\Theta(Y)$ and $\Phi(X)$ such that
the absolute value of the correlation coefficient between the 
transformed variables is maximized. That is, the so-called 
{\em maximal correlation}~\cite{Gebelein41,Hirschfeld35,Renyi70}

\beq
\label{psi}
 \Psi (X,Y) = \sup_{\Theta^\ast, \Phi^\ast} |R(\Theta^\ast(Y),\Phi^\ast(X))|
\eeq

is calculated. 
The functions $\Theta(Y)$ and $\Phi(X)$ for which the supremum is attained 
are called {\em optimal transformations}. Generalizing the concept of linear 
correlation, where the linear correlation coefficient 
$R(X,Y)$ quantifies linear dependences

\beq
Y=aX+\eta\quad(a\in R)\;,
\eeq

$\Psi (X,Y)$ quantifies nonlinear dependences of the form 

\beq
\label{thetaphi}
\Theta(Y)=\Phi(X)+\eta\;.
\eeq

Especially, if there is complete statistical dependence~\cite{Renyi70},
i.e., $Y$ is a function of $X$ or vice versa, the maximal correlation 
attains unity.
This is also true for the relation~(\ref{thetaphi}) with $\eta=0$.
Here we are mainly interested in the estimation of the optimal 
transformations for the multivariate regression problem

\beq \label{multi_dim}
   \Theta(Y)= \Phi_1(X_1) + \ldots + \Phi_m(X_m)+\eta\;.
\eeq

This is an additive model for the (not necessarily independent) input 
variables $X_1,\ldots, X_m$.
The regression functions involved can be estimated as the optimal 
transformations for the multivariate problem 
analogous to Eq.~(\ref{psi}).
To estimate these in a nonparametric way, we use the 
{\it Alternating Conditional Expectation} (ACE) algorithm~\cite{breiman85}.
This iterative  procedure
is nonparametric because the optimal transformations are
estimated by local smoothing of the data using kernel estimators.
We use a modified algorithm\footnote{
A MATLAB- and a C-program for the ACE algorithm can be obtained from the 
authors or from the web page 
{\tt http://www.fdm.uni-freiburg.de/$\sim$hv/hv.html.}
}
in which the data are rank-ordered before the optimal 
transformations are estimated.
This makes the result less sensitive to the data distribution.
We remark that optimal transformations for multiplicative combinations
of variables $\tilde{X}_1, \ldots, \tilde{X}_l$ can be estimated by forming
\beq
  X_i = h_i(\tilde{X}_1, \ldots, \tilde{X}_l)\;,
\eeq
where the $h_i$ are arbitrary functions.

With respect to the analysis of data from nonlinear dynamical 
systems, the maximal correlation and optimal transformations
have been applied to identify delay~\cite{voss97} and partial differential 
equations~\cite{vossp4}. 
In application to differential equations, time derivatives
have to be estimated from data. The effects of the noise
on the estimated optimal transformations is not yet completely understood.
On the one hand, for neglectable amounts of noise this approach has 
successfully been applied to experimental physical data of 
different origin~\cite{voss99,voss99a}, yielding also quantitatively accurate 
results.
On the other hand, if noise contamination is strong,
this method should more be used as an exploratory tool in the process of 
model selection.
To minimize the influence of the noise on the results, the 
variable with the best signal-to-noise ratio, usually the undifferentiated
time series, should be chosen as $Y$. 
In Section \ref{resultsII} we show an application to a measured time series.

\section {The data} \label{sec_data}

The time series that we will analyze in Section \ref{results}
by the methods described in Section \ref{methods} was taken
from an electric circuit in a chaotic regime. Technical details of 
the circuit are given in~\cite{rulkov92}. The data were measured at 
The Institute for Nonlinear Science of UCSD  and
are available in the scope  of the Y2K Benchmarks of Predictability 
competition at {\tt http://y2k.maths.ox.ac.uk}

The model proposed in~\cite{rulkov94,rulkov92} to describe the
circuit reads in dimensionless units

\begin{eqnarray} 
\dot{x} & = & y \nonumber\\
\dot{y} & = & -x-\delta y + z \label{ori_eqns2}\\
\dot{z} & = & \gamma(\alpha f(x) - z ) -\sigma y\;,\nonumber
\end{eqnarray}
where $x$ corresponds to a voltage, $y$ to a current and $z$ to another
voltage. The parameters correspond to combinations
of an inductivity, the resistances and the two capacitances of the circuit.
For the chosen experiment the proposed parameters are
$\alpha=15.6$, $\gamma=0.294$, $\sigma=1.52$, 
and $\delta=0.534$~\cite{rulkov94}.

The nonlinearity is given by
\beq \label {ori_nlin}
    f(x) = \left\{ \begin{array} {l l} 
  0.528 & \mbox {  if } x\le -1.2 \nonumber \\
  x(1-x^2) & \mbox {  if } -1.2 <x< 1.2 \;.\\
  -0.528 & \mbox {  if } x \ge 1.2 \nonumber
\end{array}
\right.
\eeq

The measured time series corresponds to the $x$-component
of the differential equation.

The nonparametric modelling by optimal transformations requires
a reformulation of Eq.~(\ref{ori_eqns2}) 
as a scalar higher order differential equation. 
The equivalent description reads:

\beq \label{1D_dgl}
    x^{(3)} +(\delta+\gamma) \ddot{x} +(1+\gamma \delta+\sigma)  \dot{x} 
                + \gamma (x-\alpha f(x))=0\;.
\eeq

For the sake of clarity, we define $a=\delta + \gamma$, 
$b=1+\gamma\delta + \sigma$ and $ g(x)=\gamma (x-\alpha f(x))$.
For the proposed parameters, $a=0.828$ and $b=2.677$.
Furthermore, to facilitate the comparison of the results of the different 
approaches in Section \ref{resultsIII}, based on Eq.~(\ref{1D_dgl}),
we transform the original system~(\ref{ori_eqns2}), into

\begin{eqnarray} 
\dot{x} & = & v  \nonumber\\
\dot{v} & = & w \label{trans_eqns2} \\
\dot{w} & = & -a w \,-\, b v \,-\, g(x) \nonumber
\end{eqnarray}

Visual inspection of the time series indicates that the variance of 
the observational noise is rather small. This is confirmed by
spectral analysis
which shows a low power flattening for high frequencies~\cite {lenny_privat}.
Thus, we assume that the observational noise is due to
discretization during sampling and follows a uniform
distribution:

\beq
    \eta \sim U(-0.0005, 0.0005) \quad .
\eeq

\section{Results} \label{results}

\begin{figure}
\epsfxsize=\textwidth
\epsfbox{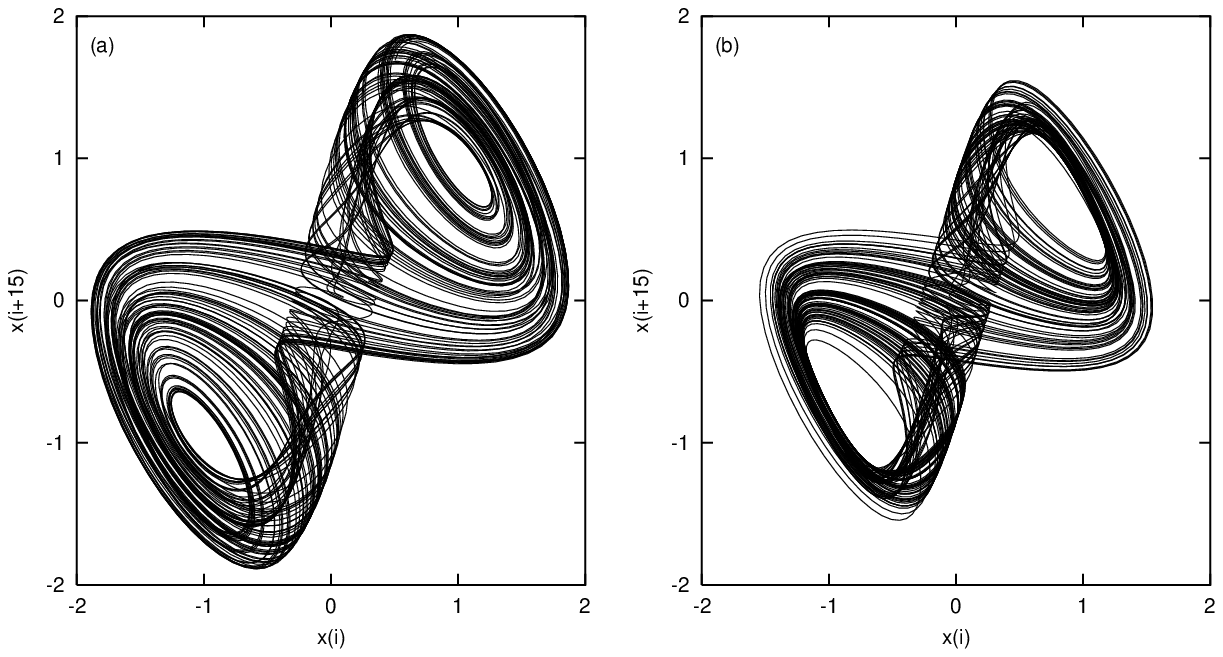} 
\caption{\label{data_simu} Reconstructed attractors. 
(a) From the measured time series. (b) From the simulated
time series based on Eq.~(\ref{ori_eqns2}) 
and parameters 
$\alpha=15.6$, $\gamma=0.294$, $\sigma=1.52$, and $\delta=0.534$. 
The delay time is 15 sampling units, corresponding to $300 \mu $s. }
\end{figure}

Figure~\ref{data_simu} shows attractor reconstructions for the
measured and a simulated time series based on Eq.~(\ref{ori_eqns2}),
and the proposed parameters given in the previous section. For the
simulated time series, the attractor is smaller, the ``holes'' in the 
two loops 
are larger and the distances between the inner edges of the loops and the
region in phase space where the trajectories change the 
loops are smaller.

\subsection {Parametric modelling I} \label{resultsI}

\begin{figure}
\epsfxsize=\textwidth
\epsfbox{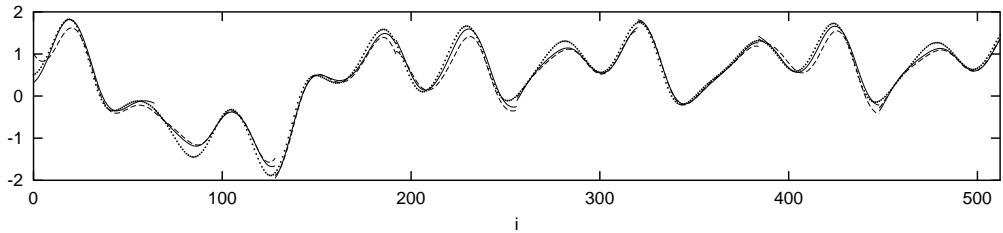}
\caption{Segment of the measured time series (dotted line),
the best fit trajectory with fixed parameters $\alpha=15.6$, $\gamma=0.294$, 
$\sigma=1.52$, and $\delta=0.534$ (broken line) and result for the best fit 
parameters $\alpha=1333$, $\gamma=0.00291$, $\sigma=1.420$, and $\delta=0.799$ 
(solid line).
The fit trajectories contain eight continuous pieces of 64 data points each.
}
\label{first_fit}
\end{figure}

To compare the measured and the simulated time series in 
the time domain we apply a restricted version of Bock's
algorithm. As outlined in Section \ref{methods_bock} the algorithm
finally yields estimates for the parameters as well as for the unobserved
dynamical state (in Eq.~(\ref{ori_eqns2}) denoted by $x,y,z$).
Here, we fix the parameters during the optimization process to the 
proposed ones and only allow the estimation of the state to be optimized.
Figure~\ref{first_fit} shows a segment of the measured time
series (dotted line) and the estimated time series conditioned
on the parameter values $\alpha=15.6$, $\gamma=0.294$, $\sigma=1.52$, 
and $\delta=0.534$ (broken line). 
We have reproduced identical results 
for numerous different choices of the initial guesses for the unobserved
components $y$ and $z$. The maximal length of the pieces applied 
in the estimation procedure that yielded a convergence was $1280\mu $s 
corresponding to 64 data points, which represents approximately 1.5 rotations
on the loops.
The mean squared error between the measured and the estimated time 
series is $2.312*10^{-2}$. 
Now we are in the situation mentioned in the Introduction, Section \ref{intro}:
The proposed model type with proposed specific parameters
is not able to reproduce the measured time series.

To decide the question if a change of the parameters
is sufficient to explain the measured data or whether
a different model type has to be chosen, we apply the full
version of Bock's algorithm. 
As initial guesses for the parameters we chose those originally proposed
and also tested different ones in the same order of magnitude.
As initial guesses for the initial values of the unobserved 
$y$, $z$-components we chose values between $-10$ and 10.
The result is given in Fig.~\ref{first_fit} (solid line).
The estimated parameters are
$\alpha =1333 \pm 101$, $\gamma =0.00291 \pm 0.00031$, 
$\sigma =1.420\pm 0.075$, and $\delta =0.799\pm0.075$.
The mean squared error between the measured and the estimated time series 
is $7.689*10^{-3}$. The mean deviation to the measured time series
is decreased, but there are still systematic discrepancies.
This can be observed by the mismatch at the minima and maxima of the time series.
Moreover, as in the first attempt, the maximal length of the pieces applied 
in the estimation procedure that yielded a convergence was 1280 $\mu $s.
As mentioned in Section \ref{methods_bock} this gives evidence that the
proposed model type is incorrect.

\subsection {Nonparametric modelling} \label{resultsII}

\begin{figure}
\epsfxsize=\textwidth
\epsfbox{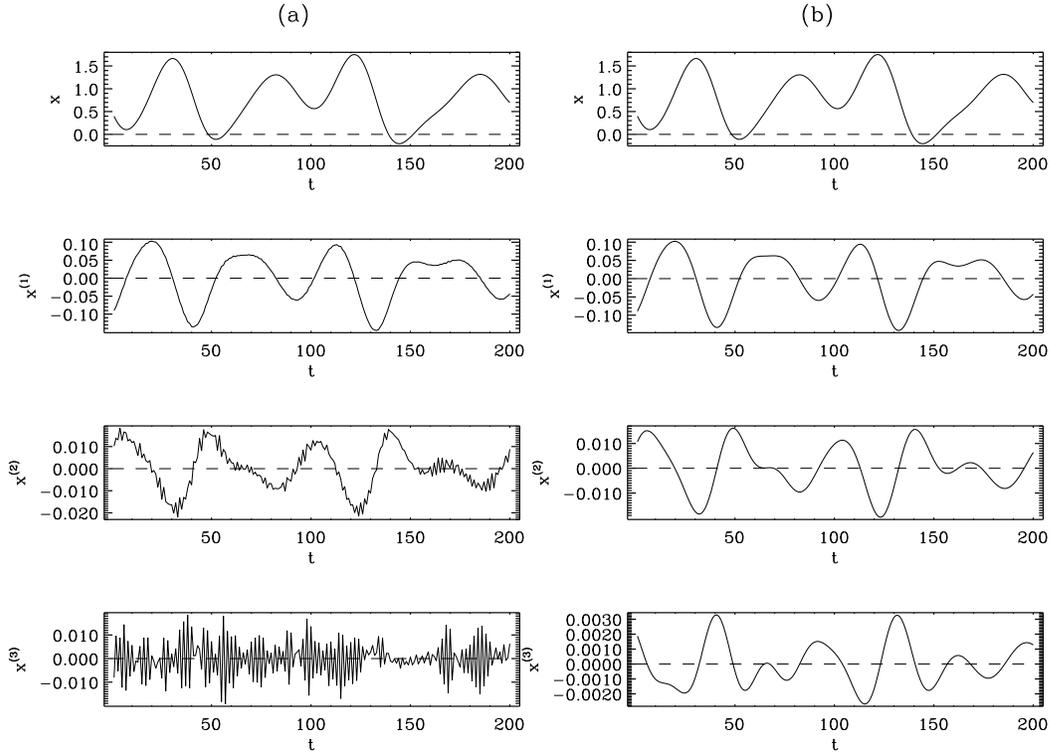}
\caption{
Piece of the time series and, from top to bottom, 
pieces of first to third time derivatives
estimated from the time series.
To obtain the highest accuracy, the derivative 
estimations are performed in frequency 
domain by multiplying the transformed data with the first to third power 
of the wave numbers, respectively, 
and subsequent back-transforming into the time domain.
(a) is a direct estimate, (b) uses a cut-off at 
10\% of the Nyquist frequency corresponding to a low-pass filter. 
The reliability of the derivative estimates can be checked roughly by 
noting that the extrema of the $i$th derivative  ($i=0,1,2$) always correspond 
to a zero in the $(i+1)$th derivatives one line below.
}
\label{3_abtl}
\end{figure}

Figure~\ref{first_fit} shows that there is a systematic deviation
of the estimated time course from the measured one even for the
best fit parameters. This calls for an alteration of the model.
Since there is only one nonlinearity in the suggested model,
see Eq.~(\ref{ori_eqns2}), we suspect that the proposed functional
form of this nonlinearity might not be correct. To obtain
a nonparametric estimate of the functional form, we apply the method
of optimal transformations, described in Section \ref{methods_opt_trans}.
Therefore, in order to make all variables accessible,
we use the equivalent one-dimensional third order system~(\ref{1D_dgl}).
As input variables for the multiple nonparametric regression 
problem~(\ref{multi_dim}) we use
\beq
Y    =   {x}, \quad
X_1  =  \dot{x} , \quad
X_2  =  \ddot{x} , \quad
X_3  =  x^{(3)}\;.
\eeq
The time derivatives entering $X_1$, $X_2$ and $X_3$ have to be estimated 
from the data. The observational noise on the data is rather small, 
see Fig.~\ref{first_fit} (solid line),
presumably only resulting from the analog-to-digital conversion.
Despite the small variance the estimated time derivatives are
rather noisy. Figure~\ref{3_abtl}a shows segments of the estimated
derivatives of first to third order.
The third derivative appears to be useless for a further analysis.
However, by a proper filtering in frequency domain, 
also third-order derivatives can be recovered
easily and with high precision, as shown in Fig.~\ref{3_abtl}b.

According to Eq.~(\ref{1D_dgl}) we expect the optimal transformations of 
$X_1$, $X_2$ and $X_3$ each to be linear.
The optimal transformation of $Y$ should turn out to be a linear combination
of the unknown nonlinearity in the circuit and a linear function.
It corresponds to $g(x)=\gamma (x-\alpha f(x)) $ in the original system.

\begin{figure}
\epsfxsize=\textwidth
\epsfbox{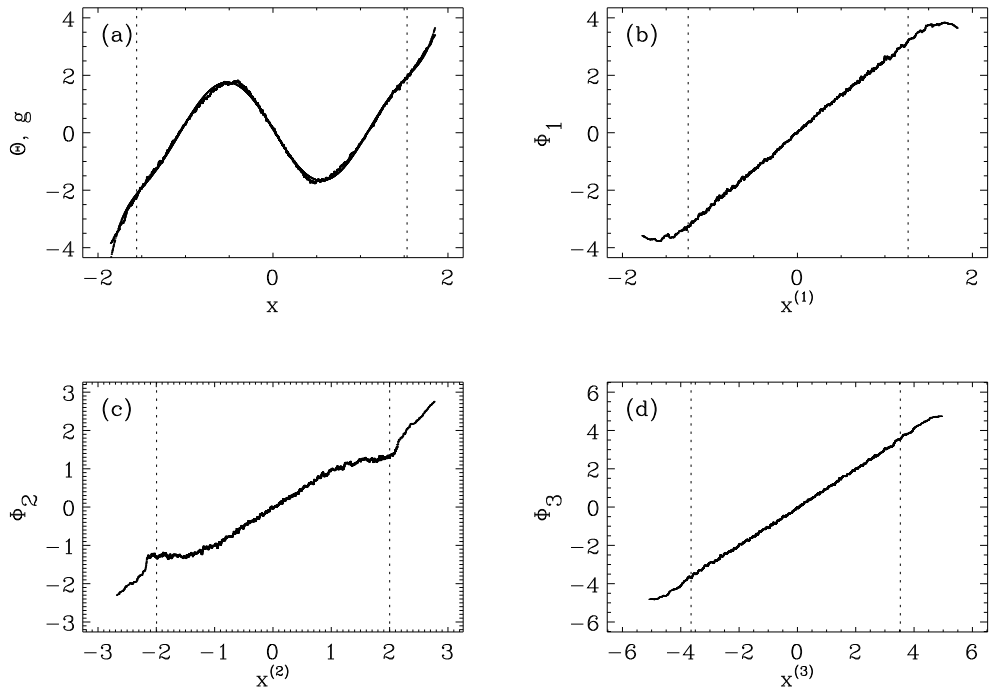}
\caption{Estimated optimal transformations for
the one-dimensional third order differential equation Eq.~(\ref{1D_dgl}). 
(a) $\Theta(Y=x)$, (b)  $\Phi_1(X_1= \dot{x}=x^{(1)})$, 
(c) $\Phi_2(X_2= \ddot{x}=x^{(2)})$, (d) $\Phi_3(X_3= x^{(3)})$.
In (a) also a fit of a 7th order polynomial is shown.
The coefficients of the nonlinearity $g(x)$ 
estimated this way are given in the text. The vertical lines indicate 
upper and lower 5 \% quantiles of the data.}
\label{opt_trafo}
\end{figure}

Applying the nonparametric regression analysis, 
we get the optimal transformations as displayed in 
Fig.~\ref{opt_trafo}.
In estimating conditional expectation values as necessary in the 
ACE algorithm, we choose a rectangular smoothing kernel
that averages over 41 data points.
The value of $\Psi(Y,X_1,X_2,X_3)$ is 0.989 indicating that $Y$ can be 
well described by the chosen variables $(X_1,X_2,X_3)$. The results
for $X_1$ and $X_3$ within the 5 to 95 \% quantiles of the data 
are well consistent with the expected linear behavior from Eq.~(\ref{1D_dgl}).
The optimal transformation for $X_2$ can still be described fairly
well by a linear function.
Linear regression yields the parameter estimates $a=0.7963$ and $b=2.5041$.
The optimal transformation for $Y=x$, Fig.~\ref{opt_trafo}(a), deviates 
qualitatively from the proposed form $g(x)=\gamma (x-\alpha f(x))$ 
which exhibits a discontinuous first derivative at $x=\pm1.2$, see 
also Fig.~\ref{ace_fit} below.
It suggests a description of the term $g(x) $ by a polynomial
\beq \label{opt_nlin}
   g(x) = \sum_{i=0}^{m} c_{i} x^{i}\;.
\eeq

We fitted polynomials of increasing order to the nonparametric 
estimate of the nonlinearity by the optimal transformation. 
An order of 7 yields a fit that
is essentially unaffected by a further increase of $m$,
see Fig.~\ref{opt_trafo}a (smooth line).
The resulting parameters are $c_0=0.1033$, $c_1=-4.9573$, 
$c_2=-0.2462$, $c_3=6.887$, $c_4=0.1744$, $c_5=-2.656$,  
$c_6=-0.040$, and $c_7=0.3652$. 

\begin{figure}
\epsfxsize=.8\textwidth
\centerline{\epsfbox{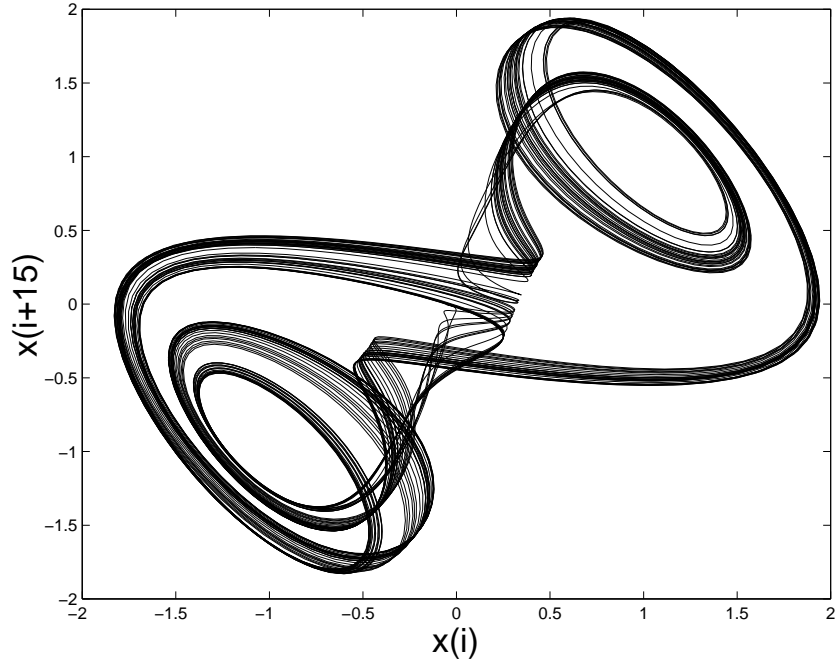}}
\caption{Reconstructed attractor based on the results for 
the nonparametric modelling by the optimal transformations.}
\label{att_maxcorr}
\end{figure}

Figure~\ref{att_maxcorr} displays the reconstructed attractor based
on a simulation of the model described by the 
estimated optimal transformations. 
The size of the attractor coincides with
the attractor reconstructed from the measured time series, but
the appearance differs. 
Since, unlike in the parametric approach, 
we have not optimized the dynamics of our model 
but performed only a nonlinear regression 
analysis of the variables $(Y,X_1,X_2,X_3)$, 
it can be expected that this result could still be improved.
This would require, however, an extensive search for optimal 
parameters in the estimation of derivatives and conditional 
expectation values in the ACE algorithm, since 
the effect of noise in these steps is not completely understood.
Rather to do that, we take this result as an exploratory approximation 
that yields useful guesses for the functional form of the nonlinearity
and the initial parameters in parametric modelling again.

\subsection {Parametric modelling II} \label{resultsIII}

In the following, to allow for a comparison
of the different approaches, we use the writing of
Eq.~(\ref{trans_eqns2}),
where the proposed nonlinearity reads $g(x)=\gamma (x-\alpha f(x))$.

The coefficients of the even-order monomials fitted to the 
optimal transformation in the previous section are rather small.
Therefore, we suspect that they are consistent with zero.
Based on the suggested form for the nonlinearity from the previous 
section, we now apply Bock's algorithm using a sum of odd monomials
up to seventh order as nonlinearity

\beq \label{neu_nlin}
   g(x) = \sum_{i=1}^{4} c_{2i-1} x^{2i-1} \quad .
\eeq

\begin{figure}
\epsfxsize=\textwidth
\epsfbox{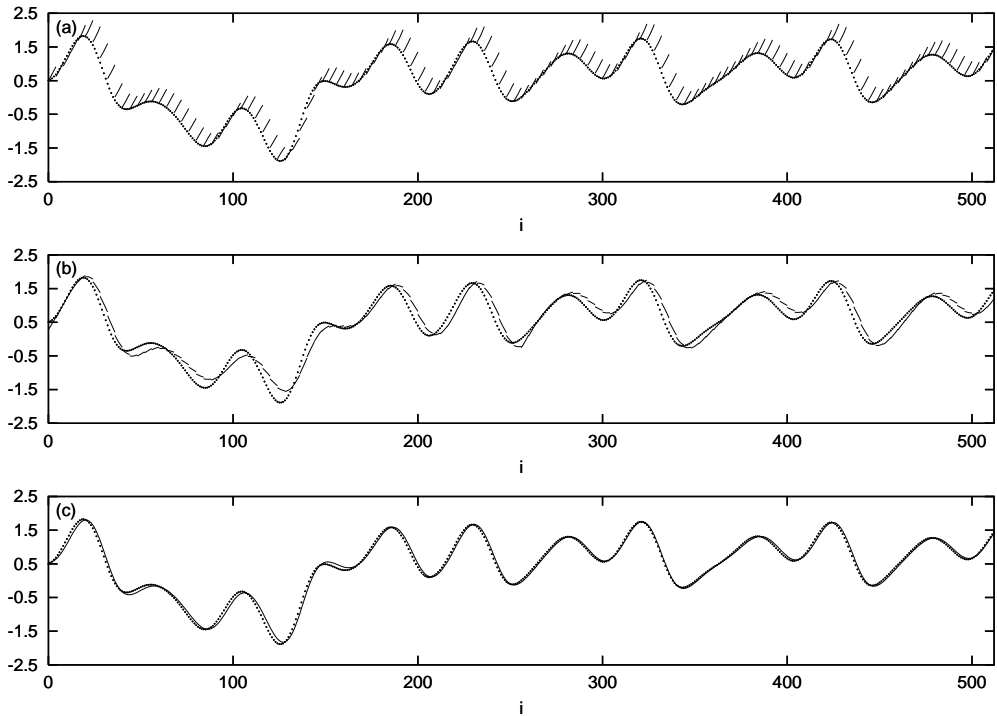}
\caption{Segment of the measured time series (dotted line)
and convergence of the multiple shooting approach based on the polynomial 
nonlinearity (solid line).
(a)~Trial trajectory for initial guesses of the parameters and initial
values, see text for details. (b) Trial trajectory after 3 iterations.
(c) Result after convergence.}
\label{second_fit}
\end{figure}

For the new model we were able to fit the parameters from pieces of
length 1024 points. 
Figure ~\ref{second_fit} shows the first half of 
one such piece of the measured time series (dotted line) and the process 
of convergence of Bock's multiple shooting algorithm (solid line).

The convergence for rather long pieces as well as the coincidence of 
the finally estimated time course of $x(t_i)$ 
and the measured data in Fig.~\ref{second_fit}c  
suggests that the polynomial nonlinearity of Eq.~(\ref{neu_nlin})
provides a better description of the underlying system than the
saturating nonlinearity of Eq.~(\ref{ori_nlin}).

As mentioned in Section~\ref{sec_data}, the small observation noise
on the data leads to unrealistic small confidence intervals for
the estimated parameters. 
To obtain a realistic impression of the variability of the 
parameter estimates, we divided the time series in 8 parts
and report the  mean and standard deviations calculated from
the results for these parts: $a=0.759 \pm 0.021$, 
$b=2.526 \pm 0.014$, $c_1=-4.56 \pm 0.16$, $c_3=6.44 \pm 0.58$, 
$c_5=-2.55 \pm 0.41$, and $c_7=0.366 \pm 0.074$.
The mean squared error is $1.581 \ast 10^{-3}$.

Figure~\ref{ace_fit} compares the proposed nonlinearity 
$g(x)=\gamma (x-\alpha f(x))$,
the optimal transformation $\Theta (Y= x)$
and the corresponding function based on the final parametric fit. 
Interestingly, the fitted polynomial closely follows the proposed
nonlinearity for small absolute values of $x$, but approaches the estimated
optimal transformation for larger values of $x$.

\begin{figure}
\epsfxsize=\textwidth
\epsfbox{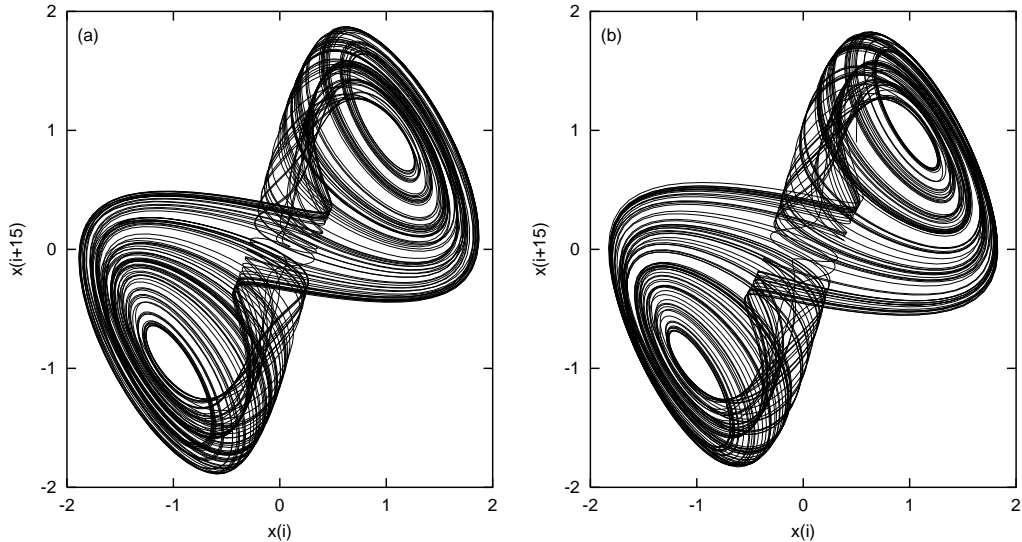}
\caption{Reconstructed attractors. 
(a) From the measured time series. (b) From the simulated
time series based on Eqs.~(\ref{trans_eqns2},\ref{neu_nlin}) 
and the best fit parameters, see text. 
The delay time is 15 sampling units, corresponding to $300 \mu $s. }
\label{data_fit}
\end{figure}

Analogous to Fig.~\ref{data_simu}, Fig.~\ref{data_fit} displays the 
reconstructed attractor based on a simulated time series using the 
polynomial Eq.~(\ref{neu_nlin}) as nonlinearity and the best fit parameters. 
The attractors are in excellent agreement.

\section{Discussion}

We proposed a three-step procedure for modelling nonlinear time series
by differential equations and exemplified this strategy on a physical
application. We chose the algorithmically more difficult modelling 
by differential equations in favor
to discrete-time difference equations because the results of the
former are usually easier interpretable in terms of the underlying
physics~\cite{timmer98d}.

Bock's multiple shooting approach to parameter estimation in
differential equations does not require to estimate time derivatives
from the data which limits the applicability of many other approaches.
Note that even for the very clean  data in our application
it is not possible to estimate the third derivative without
using some kind of low-pass filtering, see Fig.~\ref{3_abtl}.
The price to be payed in Bock's algorithm is that this
approach needs a parameterized model. 
On the other hand, nonparametric modelling by optimal transformations
does not require a specific parameterized model but is 
more susceptible to noise. 
Furthermore, it treats the problem as a case of regression, not
taking into account that the data were generated by a dynamical system.
Nevertheless, as our application has shown, it can be applied to
inspire parametric models that, again, can be checked by parametric
modelling of the dynamical system by Bock's multiple shooting algorithm.

The difference between the results for the nonparametric and the final 
parametric analysis displayed in Fig.~\ref{ace_fit} shows that a simple 
parametric fit to the nonparametric estimate of the nonlinearity as 
reported in Section \ref{resultsII} 
can be improved by the 
third step of our procedure that evaluates the predictive 
ability of the model in the time domain. 
In the discussed application our final result is extremely accurate 
considering that we used an experimental 
time series and that we can reproduce the global dynamics of this highly 
nonlinear system (Fig.~\ref{data_fit}). 
The excellent agreement between the dynamics of a measured chaotic time 
series and an estimated model is a highly nontrivial result~\cite{kantz97} 
in nonlinear modelling.

Canonically, dynamical systems are given as vector-valued first order
differential equations. 
A precondition to apply the second, nonparametric, model generating step of 
the proposed strategy is that the dynamical system can be expressed
as a scalar, higher order differential equation. 
It follows from Theorem III of Takens' seminal paper~\cite {takens81} that 
for an  $m$-dimensional system of first order differential equations
there is always an equivalent one dimensional differential equation of maximum
order $2m+1$. Unfortunately, it depends on the given system whether it
is possible to find an explicit form for the one dimensional counterpart.
Fortunately, as far as known to the authors, for all chaotic 
standard systems the one-dimensional writing is possible.
Note that in the investigated case of electronic circuits 
(due to Kirchhoff's laws~\cite{Kirchhoff45})
a huge class of nonlinear circuits can be modelled by differential
equations like Eq.~(\ref{ori_eqns2}) anyway.

It has to be emphasized that the suggested procedure of
alternating parametric modelling by Bock's algorithm
and nonparametric modelling by optimal transformations
is not a general purpose procedure in the sense of
``Equations of motion from a data series''~\cite{crutchfield87}.
The success of the discussed application depended
on prior knowledge about the system, i.e. a roughly correct
first suggestion on the structure of right hand side
of the differential equation. 
This, however, is often the case in systems where the dynamics is 
qualitatively understood, and one is interested in obtaining 
coefficients or special forms of nonlinearities involved in the dynamics.
Therefore, we assume that our approach 
will be applicable mainly for systems from physics, like
electronic circuits or lasers~\cite{weiss95,timmer00e},
engineering, e.g.~effects of nonlinear friction~\cite{moon87}, biochemistry, 
e.g.~dynamics of protein folding~\cite{creighton92},
and biophysics, e.g.~dynamics of photosynthesis~\cite{baake92b}.

\section{Acknowledgments}

We would like to thank H.I.D. Abarbanel, N. Rulkov and L. Smith
for making these interesting time series available within
the Y2K Benchmarks of Predictability Competition at
{\tt http://y2k.maths.ox.ac.uk}.

\clearpage

\begin{figure}
\epsfxsize=.8\textwidth
\centerline{\epsfbox{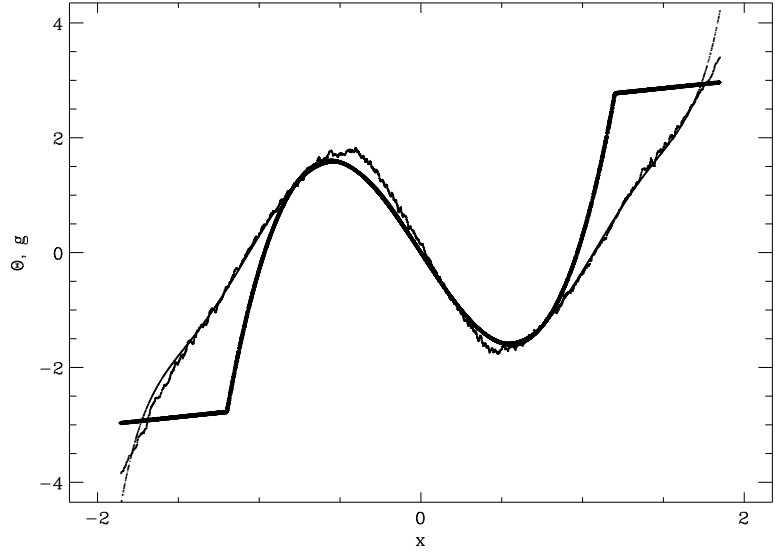}}
\caption{Comparison of the proposed piecewise differentiable model 
nonlinearity, $g(x)=\gamma (x-\alpha f(x))$ (bold line),
the nonparametrically given optimal transformation 
$\Theta(x)$ (dithered line), and 
the parametrically fitted odd polynomial (dotted line).
}
\label{ace_fit}
\end{figure}

\end {document}